\pdfoutput=1
\documentclass[11pt]{article}
\usepackage{hyperref}
\usepackage{tabularx, xcolor} 

\usepackage[final]{acl}
\usepackage{arydshln} % 표 점선 구분 위해 필요
\usepackage{times}
\usepackage{color}
\usepackage{latexsym}
\usepackage{makecell}
\usepackage{multirow}
\usepackage{amsmath}
\usepackage{subcaption}
\usepackage{amsfonts}
\usepackage{mathrsfs}
\usepackage{adjustbox} % preamble에 추가
\usepackage{graphicx}
\usepackage{threeparttable}
\usepackage{natbib}
\usepackage[T1]{fontenc}
\usepackage{booktabs}
\usepackage{multirow}
\usepackage{graphicx}
\usepackage{xcolor}
\usepackage[most]{tcolorbox}
\usepackage{courier}            % monospaced font
\newtcolorbox{PromptBox}{%
  colback=green!5!white,
  colframe=green!50!black,
  boxrule=0.8pt,
  arc=2pt,
  left=4pt, right=4pt, top=4pt, bottom=4pt,
  fontupper=\footnotesize\ttfamily,
  title=\textbf{Prompt Template},
  coltitle=black,
  titlerule style={colframe=black,linewidth=0.6pt},
  width=\columnwidth,       % ← 여기서 한쪽 열 전체 폭으로 고정
  before skip=1em,          % 위쪽 여백
  after skip=1em,           % 아래쪽 여백
}

\usepackage[utf8]{inputenc}

\usepackage{microtype}

\usepackage{inconsolata}

\usepackage{graphicx}

\title{Decoding Dense Embeddings: \\Sparse Autoencoders for Interpreting and Discretizing Dense Retrieval}
\author{Seongwan Park \\
  SungKyunKwan University \\
  Republic of Korea \\
  \texttt{waniboyy@gmail.com} \\\And
  Taeklim Kim \\
  SungKyunKwan University \\
  Republic of Korea \\
  \texttt{kimtaeklim2002@gmail.com} \\\And
  Youngjoong Ko\thanks{Corresponding author} \\
  SungKyunKwan University \\
  Republic of Korea \\
  \texttt{yjko@skku.edu} \\}

\begin{document}
\maketitle
\begin{abstract}
Despite their strong performance, Dense Passage Retrieval (DPR) models suffer from a lack of interpretability. In this work, we propose a novel interpretability framework that leverages Sparse Autoencoders (SAEs) to decompose previously uninterpretable dense embeddings from DPR models into distinct, interpretable latent concepts. We generate natural language descriptions for each latent concept, enabling human interpretations of both the dense embeddings and the query-document similarity scores of DPR models. We further introduce Concept-Level Sparse Retrieval (CL-SR), a retrieval framework that directly utilizes the extracted latent concepts as indexing units. CL-SR effectively combines the semantic expressiveness of dense embeddings with the transparency and efficiency of sparse representations. We show that CL-SR achieves high index-space and computational efficiency while maintaining robust performance across vocabulary and semantic mismatches\footnote{The code is available at \url{https://github.com/Tro-fish/Decoding-Dense-Embeddings}}.
\end{abstract}

%k 첫 줄 their 제거함. DPR과의 비교?보다는 설명가능성에 대해 강조하면 좋을 것 같다는 생각이 듦. 1)dpr은 성능이 좋지만 설명가능성이 없다. -> (브릿지 필요) -> 우리는 ~~~ 제안한다. 차라리 2) ~~이러한 시도들이 있었다? 라는 식으로 적는게 정리가 괜찮을지도. 글이 끊기는 느낌이 있음. 차라리 
%k 1) DPR은 임베딩으로 문서를 표현한다. 2) 설명가능성이 없다. 3)우리는 임베딩을 의미있는 latent로 바꾸는 것을 제안한다. with SAE 4) 기존의 임베딩과 달리 latent는 각각에 대해 의미가 존재하고 concept을 가지게 된다. 5)이는 실제로 sparse retrieval에 사용 가능하고 실험적으로 증명했다. 4) 최종적으로 latent에 대한 분석 및 해석을 통해(LLM) 각 의미를 해석가능한 텍스트로 변경하여 각각의 콘셉이 존재함을 보였다. 5)우리는 설명가능하고 빠르고 견고한 시스템 CL-SR을 제시한다. 인트로랑 뭔가 다르다 인트로 시작은 sparse인데 abstract은 dpr은 별로다 시작. 차라리 임베딩을 바꾸려는 시도와 같은 연구들로 시작하는게 어떨까. 
%k 다른방법 ) 현재 정보검색은 임베딩을 기본으로 구성된다. 크게 Dense와 sparse 방식이 있고. 이를 변형하려는 다양한 시도들이 있었다. 모두 성능 위주의 개선이였고, 설명가능성에 부합하지 않다. 우리는 SAE를 응용하여 설명가능한 임베딩을 구성하는 것을 목표로 새로운 연구를 제시한다. 3) 부터 시작 <-이런 방향으로 작성하면 참고문헌이나 서론이 풍부해지지 않을까 싶음.. 왜냐면 지금 너무 sparse dpr 두 개념을 놓고 쓴 것 같아서 연구들에 대해 조금 넣어주면 좋을 듯

%k 서론에 Sparse retrieval 얘기가 나오는게 이상하다 -> 5절로 빼기?
%k 요약에서 DPR 얘기 좀 보충하기
%k 관련 연구가 서론에 있음

\section{Introduction}

Traditional information retrieval methods have relied on exact lexical matching between query and document terms to determine document relevance \citep{Craswell2018}. Despite their efficiency and transparency, these sparse retrieval techniques suffer from \textit{vocabulary mismatch}, where a query and the relevant documents use different terms (e.g., \textit{cat} vs. \textit{kitty}), and \textit{semantic mismatch}, where the same term can refer to different concepts (e.g., \textit{bank of river} vs. \textit{bank in finance}) \citep{Gao2021COIL}.

\begin{figure*}[!t]
  \centering                      % \noindent 대신 \centering 권장
  %           왼  아래  오른  위 (단위: bp, in, cm 모두 가능)
  \includegraphics[width=\textwidth,
                   trim=0.25cm 1.6cm 0.25cm 0.3cm, clip]
                   {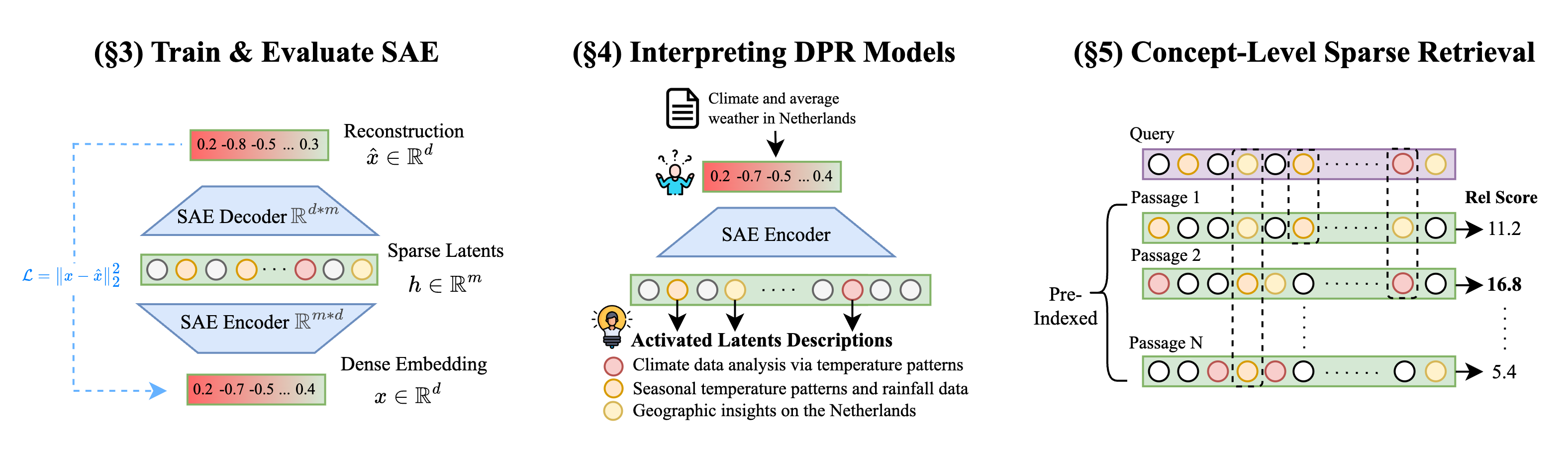}
  \caption{Overview of our method. We first train a SAE to decompose dense embeddings into latent concepts (\S\ref{sec:SAE-Training-Evaluation}). Given a query or a passage, the SAE encoder sparsely activates latent concepts, which are mapped to natural language descriptions allowing human interpretability tasks (\S\ref{sec:Interpreting_DPR}). In CL-SR, queries and passages are represented as sets of activated latent concepts (\S\ref{sec:CL-SR}). We also demonstrate its effectiveness on subsets where traditional sparse retrieval methods struggle.}
  \label{fig:overview}
  \vspace{-0.1in}
\end{figure*}

The advent of Pre-trained Language Models (PLMs) have led to the emergence of dense retrieval approaches as promising alternatives for overcoming the limitations of sparse methods \citep{Zhao2024DenseSurvey}. Dense retrieval methods embed queries and documents onto a continuous vector space by utilizing dense embeddings to represent contextualized semantics and enabling similarity computations beyond simple keyword matches. Consequently, dense retrieval effectively addresses the vocabulary and semantic mismatch issues inherent in sparse retrieval, achieving state-of-the-art (SOTA) performance across various information retrieval (IR) benchmarks \citep{Huang2024PairDistill, Xu2024BMRetriever}. However, dense retrieval suffers from a fundamental limitation: the difficulty of interpreting the dense embeddings and the ranking results. This lack of interpretability poses a significant challenge in applications where transparency and user trust in search results are critical, leading to various attempts to interpret dense retrieval models \cite{anand2022explainable}.
Recently, sparse autoencoders (SAEs) have garnered significant attention as a method to disentangle the complex semantic structures inherent in the dense embeddings of decoder-only transformer models, into distinct and interpretable conceptual units (i.e., latent concepts) \citep{bricken2023monosemanticity, templeton2024scaling, huben2024sparse}. 

In this work, we propose a novel explainable AI (XAI) framework that extends SAEs to the field of information retrieval by applying them to the embeddings of Dense Passage Retrieval (DPR) models. Our method is summarised in Figure \ref{fig:overview}.

First, we train SAEs and present qualitative measures to evaluate SAEs' ability of decomposing the semantic information embedded in the original dense embeddings into linear combinations of latent concepts. Furthermore, we qualitatively analyze if each extracted individual latent concepts hold distinct semantic meanings (Section \ref{sec:SAE-Training-Evaluation}).

We then generate natural language descriptions for each latent concept, enabling interpretation of both the semantic content of dense embeddings and the similarity computation between queries and documents. We qualitatively evaluate our framework by performing multiple human interpretability tasks (Section \ref{sec:Interpreting_DPR}).

% 이 부분 문장이 뒤에 나오는 문장이랑 겹쳐서 잘 정리해야함 --> CL-SR combines the interpretability and efficiency of sparse representations with the semantic expressiveness of dense embeddings by leveraging semantically abstracted latent concepts.

Building on this, we futher introduce Concept-Level Sparse Retrieval (CL-SR), a retrieval framework that treats each latent concept as a fundamental unit of retrieval. Unlike traditional term-based sparse methods, which are prone to vocabulary and semantic mismatches, CL-SR mitigate these limitations by leveraging semantically generalized latent concepts. CL-SR also achieves higher retrieval efficiency with fewer matching units compared to term-based retrieval (Section \ref{sec:CL-SR}).

\section{Related Work}
\subsection{Interpretability in Dense Passage Retrieval}
Dense passage retrieval employs PLM-based encoders to embed queries and passages\footnote{In this paper, the terms “passage” and “document” are used interchangeably.} into dense vectors—typically obtained from the [CLS] token representation or mean pooling over token embeddings—and measures relevance via dot product or cosine similarity between these embeddings, exhibiting superior performance across a variety of IR tasks. However, DPR inherently suffers from a lack of interpretability due to the implicit semantics encoded within the uninterpretable dense embedding space. To address this shortcoming, various attempts have been made to interpret the inner workings of DPR. \citet{volske2021towards}, analyzes the ranking determinants between document pairs based on axioms defined in traditional IR theory—e.g., term frequency, document length, semantic similarity, and term proximity—by formalizing each axiom for rank comparison and training a simple explanatory model that approximates DPR’s ranking. A different approach, \citet{llordes2023explain} generate an "equivalent query" through discrete state-space search that reproduces DPR’s ranking results under a sparse retrieval model, thereby explaining the semantic information used by DPR at the term level. However, while these approaches provide a proxy-based interpretation that approximates neural behavior through input manipulation and external alignment, the internal representational structure of the DPR model remains unexplored.

% ---------- Table: Reconstruction & Retrieval Performance with Spearman + Std (scaled) ----------
\begin{table*}[t]
\centering
\small
\newcommand{\ratio}[1]{\scalebox{0.7}{\((\times #1)\)}}
\newcommand{\std}[1]{\scalebox{0.7}{\(\pm #1\)}}
\begin{adjustbox}{width=\textwidth}
\begin{tabular}{l c cc ccc ccc}
\toprule
\multirow{2}{*}{\textbf{Model}} & \multirow{2}{*}{\textbf{NMSE}} &
\multicolumn{2}{c}{\textbf{MSMARCO Dev}} &
\multicolumn{3}{c}{\textbf{TREC DL 2019}} &
\multicolumn{3}{c}{\textbf{TREC DL 2020}} \\
\cmidrule(lr){3-4}\cmidrule(lr){5-7}\cmidrule(lr){8-10}
& & \textbf{MRR@10} & \textbf{Recall@1k} & \textbf{NDCG@10} & \textbf{Recall@1k} & \textbf{Spearman} & \textbf{NDCG@10} & \textbf{Recall@1k} & \textbf{Spearman} \\
\midrule
Baseline (SimLM)                     & --     & 0.411 & 0.986 & 0.714 & 0.767 & -- & 0.697 & 0.772 & -- \\
\textbf{Reconstructed (k=32)}  & 0.1903 & 0.328 \ratio{0.80} & 0.964 \ratio{0.98} & 0.589 \ratio{0.83} & 0.666 \ratio{0.87} & 0.928 \std{7.7E\text{-}3} & 0.582 \ratio{0.84} & 0.681 \ratio{0.88} & 0.927 \std{8.1E\text{-}3} \\
\textbf{Reconstructed (k=48)}  & 0.1643 & 0.338 \ratio{0.83} & 0.968 \ratio{0.98} & 0.624 \ratio{0.87} & 0.686 \ratio{0.89} & 0.936 \std{6.7E\text{-}3} & 0.613 \ratio{0.88} & 0.704 \ratio{0.92} & 0.933 \std{7.2E\text{-}3} \\
\textbf{Reconstructed (k=64)}  & 0.1458 & 0.347 \ratio{0.84} & 0.973 \ratio{0.98} & 0.640 \ratio{0.90} & 0.692 \ratio{0.90} & 0.949 \std{6.2E\text{-}3} & 0.608 \ratio{0.87} & 0.718 \ratio{0.93} & 0.948 \std{6.8E\text{-}3} \\
\textbf{Reconstructed (k=128)} & 0.1069 & \textbf{0.371 \ratio{0.90}} & \textbf{0.980 \ratio{0.99}} & \textbf{0.664 \ratio{0.93}} & \textbf{0.726 \ratio{0.95}} & \textbf{0.959 \std{5.2E\text{-}3}} & \textbf{0.629 \ratio{0.90}} & \textbf{0.734 \ratio{0.95}} & \textbf{0.956 \std{5.4E\text{-}3}} \\
\bottomrule
\end{tabular}
\end{adjustbox}
\caption{Quantitative evaluation of the reconstruction capability of SAEs trained on SimLM embeddings. Results include normalized mean squared error (NMSE) and preservation of retrieval performance on MSMARCO Dev and TREC DL 2019/2020. Spearman’s correlation is averaged across TREC DL 2019 and 2020, with variance also reported ($p < 0.05$).}

\label{tab:recon_performance}
\vskip -.16in
\end{table*}
% -------------------------------------------------------------------
\begin{table}[t]
\centering
\resizebox{\columnwidth}{!}{
\begin{tabular}{lcccc}
\toprule
 & \textbf{SAE k=32} & \textbf{SAE k=48} & \textbf{SAE k=64} & \textbf{SAE k=128} \\
\midrule
\textbf{Accuracy} & 0.859 & 0.829 & 0.808 & 0.764 \\
\bottomrule
\end{tabular}
}
\caption{Results of latent intrusion test using MSMARCO passages. We report accuracy over different numbers of activated latents $k$.}
\label{tab:latent_intrusion}
\vskip -.16in
\end{table}
\subsection{Sparse Autoencoders}
\label{sec:sae_train_evaluation}
The "superposition hypothesis" suggests that neural networks are able to encode more features than their available dimensions, exploiting sparsity of feature activation \citep{elhage2022superposition}. This results in superposed representations that are difficult to interpret directly due to polysemanticity. Recent work applies a set of methods called sparse coding or sparse dictionary learning to identify underlying true features actually used by neural networks \citep{sharkey2025open}. One of the simplest and widely explored methods is SAE. SAE is a single-layer feedforward autoencoder with a hidden layer larger than the input dimension, incorporating a sparsity constraint to ensure that only a small subset of hidden neurons (i.e., latent concepts) activate for any given input. 
Recent studies have empirically demonstrated that SAEs can effectively extract interpretable features from the activations of decoder-only large language models(LLMs) \citep{huben2024sparse, marks2025sparse}. While there have been attempts to apply SAEs to dense embeddings generated by encoder-based models \citep{ye2024steering, kang-etal-2025-interpret}, these efforts have largely focused on the interpretation of the dense embeddings themselves, leaving the interpretability of retrieval results and their implications in IR tasks underexplored. 

\section{SAE Training and Evaluation for DPR Model Interpretation}
\label{sec:SAE-Training-Evaluation}
Interpreting a DPR model through SAE latent concepts requires both an effective training of the SAE and a verification that its reconstructed embeddings faithfully preserve the information from the original embeddings. Additionally, it is essential to evaluate whether each latent concept extracted by the SAE represents clear semantic concept. In this section, we outline the training and evaluation procedures for achieving these objectives.

\subsection{Training Sparse Autoencoder}
\label{sec:Training-Sparse-Autoencoder}
A SAE is optimized to reconstruct the input vector $h \in \mathbb{R}^d$ by learning a sparse latent representation $z \in \mathbb{R}^m(m>>d)$ with sparsity constraint $L_0(z)=k (k<<d)$. Formally, a SAE consists of
\begin{align}
\text{Encoder:} \quad z(h) &= \sigma(W_{\text{enc}} h + b_{\text{enc}}) \label{eq:encoder} \\
\text{Decoder:} \quad \hat{h}(z) &= z W_{\text{dec}} + b_{\text{dec}} \label{eq:decoder}
\end{align}

\noindent where $W_{enc} \in \mathbb{R}^{m*d}$, $b_{enc} \in \mathbb{R}^m$, $W_{dec} \in \mathbb{R}^{d*m}$, and $b_{dec} \in \mathbb{R}^d$ are the parameters of encoder and decoder respectively and $\sigma(\cdot)$ is the activation function. Although there are various SAE variants depending on the activation and sparsity mechanisms, we employ widely used BatchTopK SAE \citep{bussmann2024batchtopk}, where the model is trained to minimize the following loss: 
\begin{align}
L(h) &= \left\| h - \hat{h}(z(h)) \right\|_2^2 + \lambda L_{\text{aux}} , \label{eq:sae_loss} \\
z(h) &= \operatorname{BatchTopK}(W_{\text{enc}} h + b_{\text{enc}}). \label{eq:batchtopk}
\end{align}

\noindent The BatchTopK activation function masks latents whose activation values are not in the top n * k to 0 across a batch of n samples, allowing flexible allocation of the number of latents for each sample in a single batch and $L_{aux}$ is an auxiliary loss, used to prevent dead latents. At inference time, latents with activation larger than mean of top k-th activation over the whole datapoints are considered "activated" and all other activations are zeroed out resulting in a sparse m dimensional vector $z(h)$. Training the SAE requires dense embeddings generated by a pretrained DPR model. In this work, we adopt SimLM \citep{wang2023simlm} as our target model to interpret and use it to embed approximately 8.8 million passages and 0.5 million train queries from the MSMARCO passage retrieval dataset \citep{bajaj2018msmarco} into dense vectors. The SAE is then trained to reconstruct these dense embeddings\footnote{We show generalizability of our framework across target DPR models and to unseen datasets in Appendix \ref{sec:appendix_generalization}.}. We set the hyperparameter m = 32 * d and experiments with k of {32, 48, 64, 128}. We detail the training setup and dataset in Appendix \ref{sec:training_details}. 

\subsection{Evaluating Sparse Autoencoder}
\label{sec:evaluate_sae}

\begin{comment}
To evaluate whether the set of latent concepts extracted by the SAE preserves the semantic structure of the DPR model and can be decomposed into human-interpretable units, we formulate the following research questions:
\begin{itemize}
    \item \textbf{RQ1}: To what extent can the latent concepts preserve the essential semantic information embedded in the original dense DPR representations? 
    
    \item \textbf{RQ2}: Do the individual latent concepts correspond to semantically coherent and interpretable units of meaning? 
\end{itemize}
First, in relation to RQ1, we evaluate the quality of SAE reconstruction based on three complementary criteria:
\end{comment}
Now we propose to quantitatively evaluate the SAE
based on following criteria.

\begin{enumerate}
\item{\textbf{Vector-level Reconstruction Fidelity}: We measure the normalized mean squared error (NMSE) between the original DPR embeddings and the reconstructed embeddings. Specifically, NMSE is calculated by dividing the raw MSE by the baseline reconstruction error of always predicting the mean activation.}

\item{\textbf{Preservation of IR Performance}: Previous works on SAEs have measured language modeling performance change as a measure of SAE reconstruction fidelity \citep{rajamanoharan2024jumpingaheadimprovingreconstruction,gao2025scaling}. We evaluate how much of the downstream dense retrieval performance is maintained by conducting dense retrieval with the reconstructed embeddings instead.}

\item{\textbf{Ranking Result Reconstruction Fidelity}: Beyond retrieval performance, we examine how faithfully the reconstructed embeddings retain the detailed ranking order of retrieved documents. We report Spearman’s correlation between the two ranked lists obtained from the target model and reconstructed embeddings.}
\end{enumerate}

For evaluation, we used the MSMARCO Dev and TREC Deep Learning (DL) Track 2019/2020 datasets \citep{craswell2020trec2019, craswell2021trec2020}. Table \ref{tab:recon_performance} shows a trade-off between the reconstruction quality and the degree of sparsity of SAEs. This trend can be observed with all three criteria we adapted to measure reconstruction fidelity of SAEs in IR settings and is consistent with prior studies on sparse autoencoders on decoder models confirming soundness of our measures.

%Second, in relation to RQ2,
Additionally, since our goal is to interpret the
DPR model via individual latent concepts, we conduct experiments to verify whether each latent indeed represents an interpretable semantic concept. Inspired by the “word intrusion test” \citep{NIPS2009_f92586a2}, we perform a latent intrusion test, where we collect 9 passages from MSMARCO corpus that most strongly activate a given latent plus 1 randomly chosen “intruder” passage that does not activate the latent. We then use a powerful LLM (GPT-4.1 mini) to identify the outlier. Table \ref{tab:latent_intrusion} shows that, though individual latent's quality degrades as sparsity decreases, SAEs are still effective at disentangling semantic structures within dense embeddings of DPR models into latent concepts that retain distinct, meaningful information.

\section{Interpreting DPR Models Through Latent Descriptions}
\label{sec:Interpreting_DPR}
Building on the latent concepts from the trained SAE, in the following sections, we first generate natural language description for each latent concept (§\ref{sec:Description-Generation}) and then we demonstrate the utility of these descriptions in two downstream human interpretability tasks\footnote{We lay out detailed experimental settings in Appendix \ref{sec:experimental_details}.}: understanding the semantic structure of DPR’s dense embeddings (§\ref{sec:Interpreting-Dense-Embeddings}) and simulating the model’s ranking process between query and documents (§\ref{sec:Interpreting-Ranking-result}).

\subsection{Generate automatic descriptions for each latents} 
\label{sec:Description-Generation}
\begin{figure}
 \includegraphics[width=\columnwidth]{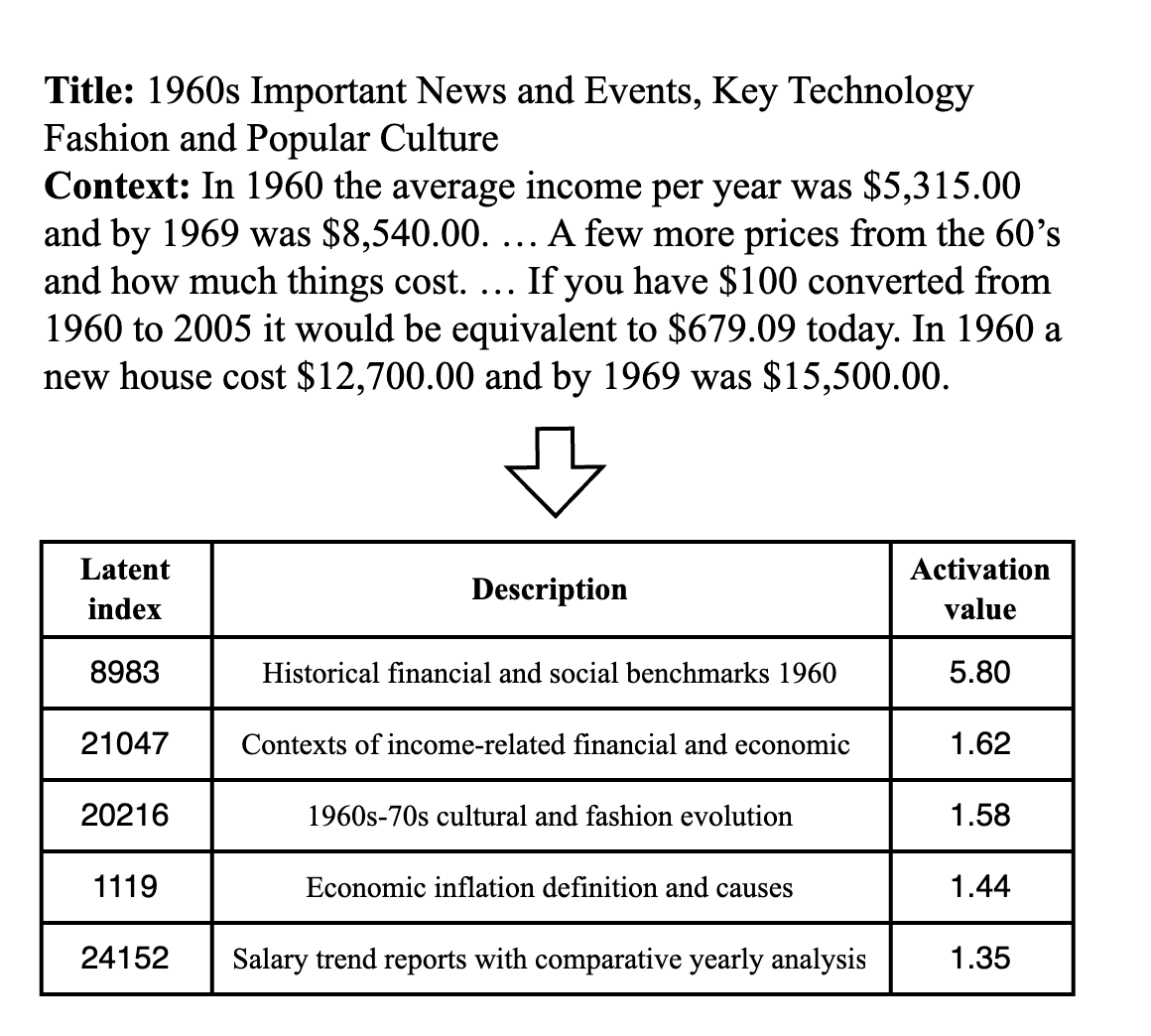}
 \caption{Top-5 most activated latent concepts extracted from the dense embedding of a passage discussing economic benchmarks and cultural context in the 1960s. Reported activation value is weighted by IDF. 
}
 \label{fig:example_latent_description}
\end{figure}

 We generate natural language descriptions enabling humans to intuitively understand the semantic meaning of each latent concept. Specifically, for each latent concept in the SAE k=32 trained in Section \ref{sec:SAE-Training-Evaluation} (which showed the best performance on latent intrusion test), we collect MSMARCO passages that most activate the corresponding latent concept and instruct a LLM to summarize the common themes, concepts, or characteristics shared across these passages. Figure \ref{fig:example_latent_description} shows how a decomposition of a document into generated latent descriptions looks like. 
 
\subsection{Interpreting Dense Embeddings via Latent Descriptions}
\label{sec:Interpreting-Dense-Embeddings}
In this section, we evaluate whether latent descriptions can be used to interpret the semantic information embedded within DPR model’s dense embeddings. To this end, we decompose a passage’s dense embedding into latent components and assess whether the corresponding latent descriptions enable accurate identification of the original passage. Specifically, each human annotator is presented with one target passage, nine randomly sampled distractor passages, and a set of latent concepts extracted from the target with their activation strengths. Annotators are then asked to identify the target passage based on the provided descriptions and activation strength.

Since the latent concepts used in this study are obtained via unsupervised training of SAE reconstructing DPR model embeddings, their activation frequency across passages varies significantly (Figure \ref{fig:latent_frequency_distribution}). In particular, high frequency latents tend to be abstract and less interpretable. To address this skewed distribution and to better capture the semantic importance of each latent, we adjust each latent’s activation strength with its Inverse Document Frequency (IDF), as defined in Eq.~\ref{eq:latent_idf}. We provide illustrative examples comparing latent descriptions with and without IDF weighting in Appendix~\ref{sec:APPENIX_interpretability_case_study}.

We randomly sampled 600 passages from the MSMARCO corpus and evaluated the embedding interpretability task and report the accuracy in Table \ref{tab:interpretability}. The high accuracy (0.943) confirms that latent descriptions effectively reveal the semantic content encoded in the DPR model’s dense representations.

\begin{table}[t]
\centering
\small
\begingroup
  \setlength{\tabcolsep}{2pt}
  \begin{tabularx}{\columnwidth}{@{} X c @{}}
    \toprule
    \textbf{Task} & \textbf{Accuracy} \\
    \midrule
    \multicolumn{2}{@{}c@{}}{\textit{Embedding Interpretability}} \\
    \cmidrule(lr){1-2}
    Passage identification & 0.943 \\
    \midrule
    \multicolumn{2}{@{}c@{}}{\textit{Ranking Interpretability}} \\
    \cmidrule(lr){1-2}
    R@P vs R@P      & 0.903 \\
    R@P vs NR@P    & 0.938 \\
    R@N vs NR@P    & 0.921 \\
    \bottomrule
  \end{tabularx}
\endgroup

\caption{Human annotator accuracy on interpretability tasks: Embedding Interpretability (§~\ref{sec:Interpreting-Dense-Embeddings}) and Ranking Interpretability (§~\ref{sec:Interpreting-Ranking-result}). (Note: R@P = Retrieved Positive; NR@P = Not Retrieved Positive; R@N = Retrieved Negative.)}
\label{tab:interpretability}
\end{table}

\subsection{Interpreting Ranking result via Latent Descriptions}
\label{sec:Interpreting-Ranking-result}
We evaluate the interpretability of the DPR model’s similarity scoring process through latent descriptions. Specifically, we assess whether humans can simulate the ranking behavior of the DPR model. We provide human annotators two candidate passages for a given query, along with the latent descriptions and IDF-adjusted activation values extracted from each passage and query. Then they are tasked to choose from which the two documents the DPR model would have ranked higher.

We evaluate the model simulatability across three specific settings of interest:
\begin{enumerate}
\item{Retrieved Positive \textbf{vs} Retrieved Positive.}
\item{Retrieved Positive \textbf{vs} Not Retrieved Positive.}
\item{Retrieved Negative \textbf{vs} Not Retrieved Positive.}
\end{enumerate}
The experiments were conducted using queries from the TREC-DL 2019 and 2020 datasets. We define documents ranked within the top 1000 as "retrieved", and those ranked lower as "not retrieved". For each query, the official gold document is considered "positive", while all others are considered "negative"

As shown in Table~\ref{tab:interpretability}, human annotators achieve high accuracy in each settings of interest showing accuracy higher than 0.9, confirming that explanations from our framework indeed help human annotators to
simulate model predictions more accurately.

\section{Concept-Level Sparse Retrieval}
\label{sec:CL-SR}

Now we extend the sparse autoencoder framework to the sparse retrieval domain. We propose Concept-Level Sparse Retrieval (CL‑SR), a novel approach that treats each latent concept as a fundamental unit of retrieval. By replacing lexical terms with these semantically coherent and generalized latent concepts, CL-SR enables inverted index retrieval from any dense retrieval model.  CL-SR offers \textbf{Semantic generalization} and \textbf{Computational efficiency}. We further analyse these qualities in Section~\ref{sec:Retrieval-Effectiveness}, ~\ref{sec:Robustness-Analysis}.
\begin{table*}[t]
\centering
\small
\begin{threeparttable}
\begin{adjustbox}{width=\textwidth}
\begin{tabular}{l cc cc cccc}
\toprule
\multirow{2}{*}{\textbf{Model}} &
\multicolumn{2}{c}{\textbf{MSMARCO dev}} &
\multicolumn{1}{c}{\textbf{TREC DL 2019}} &
\multicolumn{1}{c}{\textbf{TREC DL 2020}} &
\multicolumn{4}{c}{\textbf{Efficiency}} \\
\cmidrule(lr){2-3} \cmidrule(lr){4-4} \cmidrule(lr){5-5} \cmidrule(lr){6-9}
& \textbf{MRR@10} & \textbf{Recall@1k} & \textbf{NDCG@10} & \textbf{NDCG@10} & \textbf{FLOPs} & \textbf{Avg. D Len} & \textbf{Storage (GB)} & \textbf{Vocab Size} \\
\midrule
\multicolumn{9}{c}{\textbf{(Unsupervised) Sparse Retrieval}} \\
\cmidrule(lr){1-9}
BM25        & 0.183 & 0.853 & 0.506 & 0.478 & 0.13 & 39.41  & 0.67 & 2,660,824 \\
RM3         & 0.165 & 0.870 & 0.522 & 0.489 & --                       & 39.41  & 0.67 & 2,660,824 \\
docT5query  & 0.277 & 0.947 & 0.626 & 0.607 & 0.94 & 756.62 & 0.98 & 2,660,824 \\
\midrule
\multicolumn{9}{c}{\textbf{(Neural) Sparse Retrieval}} \\
\cmidrule(lr){1-9}
query2doc   & 0.214 & 0.918 & 0.635 & 0.582 & 0.87 & 39.41  & 0.67 & 2,660,824 \\
DeepImpact  & 0.327 & 0.947 & 0.657 & 0.603 & --                       & 71.61  & 1.40 & 3,514,102 \\
uniCOIL     & 0.315 & 0.924 & 0.643 & 0.652 & 1.03 & 67.96  & 1.30 &   30,522 \\
SPLADE-max    & 0.340 & 0.965 & 0.683 & \textbf{0.671} & 1.35 & 91.50 & 2.60 &   30,522 \\
\midrule
\textbf{CL-SR (Efficient)} & 0.343 & 0.954 & 0.643 & 0.593 & \textbf{0.11} & \textbf{22.95} & \textbf{0.57} & \textbf{11,709} \\
\textbf{CL-SR (Max)}       & \textbf{0.368} & \textbf{0.969} & \textbf{0.686} & 0.634 & 0.74 & 64.73  & 1.27 & 18,679    \\
\bottomrule
\end{tabular}
\end{adjustbox}
\caption{
Comparison of retrieval performance and efficiency across unsupervised, neural, and our CL‑SR models on MSMARCO dev, TREC DL 2019, and TREC DL 2020. FLOPs measured by following the settings of \citep{formal2021splade}. 
query2doc performance copied from \citep{wang2023query2doc}.
All other metrics measured using Pyserini \citep{lin2021pyserini}.
}
\label{tab:retrieval_efficiency}

\end{threeparttable}
\end{table*}

In CL-SR, the scoring between a query and a passage is computed using the following formulation (Eq. (\ref{eq:latent_bm25})).

\begin{align}
  s(q, d) &= \sum_{i \in q \cap d} f_d(q, i) \cdot f_d(d, i) \cdot \mathrm{idf}(i) \label{eq:latent_bm25}
\end{align}
\vspace{-2mm}
where
\begin{align}
   & \mathrm{idf}(i) = \log \frac{|D|}{1 + \left|\left\{d \in D \mid {z(h)}_{d,i} > 0\right\}\right|}
\label{eq:latent_idf}
\end{align}
\vspace{-2mm}
\begin{align*}
    & f_q(q, i) = \frac{{z(h)}_{q,i} (1 + k_2)}{{z(h)}_{q,i} + k_2}, \\
    & \scalebox{1.1}{$f_d(d, i) = \frac{{z(h)}_{d,i} (1 + k_1)}{{z(h)}_{d,i} + k_1 \left(1 - b + b \frac{\|{z(h)}_d\|_1}{\sum_{d \in D} \|{z(h)}_d\|_1 / |D|}\right)}.$}
\end{align*}

This equation replaces the traditional BM25 term frequency with latent activation values and redefines document length normalization based on the total activation in latent space. Here, $z(h)_{q,i}$ and $z(h)_{d,i}$ denote the activation values of latent concept $i$ in the query and passage, respectively, and D denotes the entire set of documents. 

As discussed in Section~\ref{sec:Interpreting-Dense-Embeddings}, high-frequency latents tend to contribute less meaningfully to information retrieval. To mitigate their influence, we apply IDF weighting\footnote{The impact of IDF weighting on CL-SR is provided in the Appendix \ref{sec:APPENIX_IDF_CL_SR_performance}}.\\

\noindent\textbf{Implementation and evaluation setup}\\
We conducted retrieval experiments using SAEs trained in section \ref{sec:Training-Sparse-Autoencoder}. Among them, we utilze two configurations for comparison: k=32 (Efficient) and k=128 (Max). The Efficient model is optimized for computational efficiency whereas the Max model prioritizes retrieval accuracy by allowing a greater number of latent concepts to be used as document identifiers. 

Similar to traditional sparse retrieval methods, CL‑SR allows for the construction of an inverted index in advance, based on indices of latent concepts activated from the each document in the collection. To reduce storage and retrieval overhead, we index only a fixed number of highly activated latent concepts per passage by setting maximum number of allowed latents for each passage rather than indexing all the activated latents. Detailed hyperparameter settings and results for other SAE variants are provided in the Appendix \ref{sec:APPENDIX_Latent_count}. 

At retrieval time, only the query embedding is projected into the latent space, and its sparse latent activations are directly used to compute document rankings using the Eq.\eqref{eq:latent_bm25}. To assess retrieval efficiency, we measure: (1) FLOPs, defined as the expected number of floating-point operations per query–document pair; (2) the average number of activated latents per passage (analogous to number of tokens in term-based retrieval); and (3) index storage size. Specifically, FLOPs are computed as
\(\mathbb{E}_{q,d} \left[ \sum_{j \in V} p_j^{(q)} \cdot p_j^{(d)} \right]\), 
where \(V\) denotes the vocabulary, and \(p_j\) is the activation probabilities for token \(j\) in document \(d\) and query \(q\) respectively. Following prior work \citep{formal2021splade} this metric is computed over a set of approximately 100k queries, on the MSMARCO collection.
For evaluation, we use the same datasets as in Section~\ref{sec:evaluate_sae}\\\\
\noindent\textbf{Baselines}\\
We categorize the baseline models into two major groups: Unsupervised Sparse Retrieval and Neural Sparse Retrieval. The Unsupervised group includes classical term frequency–based methods such as BM25, as well as its extensions via query expansion (RM3) \citep{lavrenko2001relevance} and passage expansion (docT5query) \citep{nogueira2019docTTTTTquery}. These methods do not involve any neural network at inference time. In contrast, the Neural Sparse Retrieval group employs neural networks—typically PLMs—at inference time to dynamically reweight or expand query and document terms. Specifically, we include query2doc \citep{wang2023query2doc}, DeepImpact \citep{mallia2021learning}, uniCOIL (no expansion) \citep{lin2021few}, and SPLADE v2 (max) \citep{formal2021spladev2} as representative neural approaches.

While more recent variants such as DistilSPLADE and SPLADE v3 \citep{lassance2024spladev3} achieve stronger performance through heavy distillation and advanced training techniques (e.g., ensemble teacher re-rankers, model-specific hard negative sampling), these enhancements are orthogonal to the focus of our work. To ensure a fair comparison, we report non-distilled SPLADE v2-max as a baseline.

\subsection{Retrieval Effectiveness and Efficiency}
\label{sec:Retrieval-Effectiveness}
As shown in Table {\ref{tab:retrieval_efficiency}, the CL‑SR framework demonstrates retrieval accuracy on par with other neural sparse retrieval baselines while achieving superior computational efficiency. On MSMARCO Dev, SAE-Max achieves the highest performance with MRR@10 of 0.368 and Recall@1k of 0.969. SAE-Efficient matches the performance of SPLADE v2 with an MRR@10 of 0.343, but requires only 0.11 FLOPs per query and 0.57 GB of index storage. We believe this efficiency arises from representing documents with a compact set of semantically abstracted latent concepts, which substantially reduces the cost of query-document matching.

On TREC-DL 2019/2020,  models based on exact lexical matching are more favorable since most queries are long-tailed and entity-centric \citep{wang2023query2doc}. Nevertheless, SAE-Max still achieves competitive performance with nDCG@10 scores of 0.686 (2019) and 0.634 (2020).

Figure~\ref{fig:performance_vs_flops} illustrates the trade-off between retrieval effectiveness (MRR@10) and computational cost (FLOPs) on the MS MARCO Dev set. The performance of CL-SR shows diminishing returns as the maximum number of allowed latent concepts per passage increases. To balance efficiency and effectiveness, we configure CL-SR Efficient and CL-SR Max to index up to 24 and 65 latent concepts per passage, respectively, resulting in average document lengths (i.e., number of active latents) of 22.95 and 64.73.

\begin{figure}
 \includegraphics[width=\columnwidth]{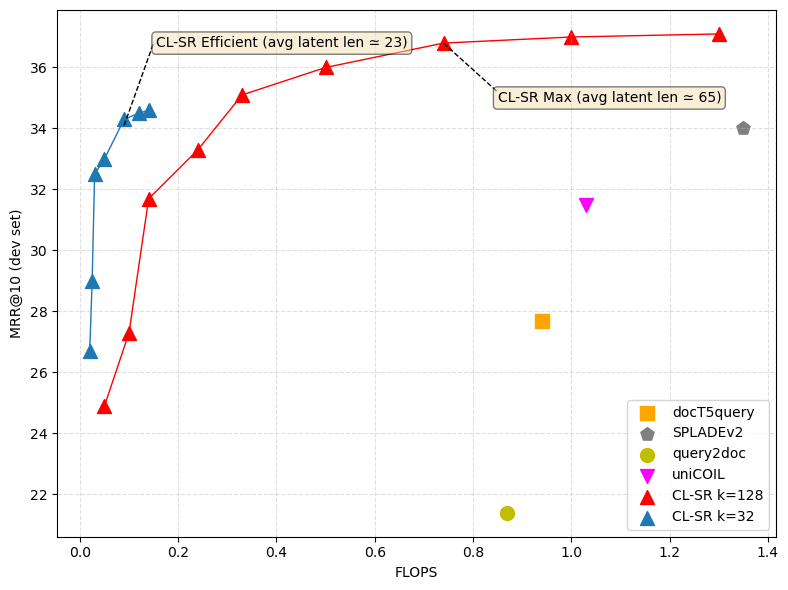}
 \caption{Performance vs FLOPs for CL-SR with varying number of latent concepts used for each document when indexing. 
}
 \label{fig:performance_vs_flops}
\end{figure}
\vspace{-2mm}
\begin{table}[t]
\centering
\small
\newcommand{\drop}[1]{\scalebox{0.7}{\textcolor{gray}{#1}}}
\begin{tabular}{lcc}
\toprule
\textbf{Models}            & \textbf{MRR@10}      & \textbf{Retrieval Type}   \\
\midrule
BM25                       & 0.0   \drop{(-100\%)}     & \multirow{2}{*}{Unsupervised} \\
docT5query                 & 0.052 \drop{(-80.8\%)}   &                             \\
\cdashline{1-3}
\\[-0.8em]
DeepImpact                 & 0.094 \drop{(-71.2\%)}   & \multirow{4}{*}{Neural Sparse}     \\
SPLADEv2                 & 0.106 \drop{(-68.9\%)}   &                             \\
CL-SR (Efficient)          & 0.124 \drop{(-63.7\%)}   &                             \\
CL-SR (Max)                & \textbf{0.143 \drop{(-61.1\%)}}   &                             \\
[0.2em]
\cdashline{1-3}
\\[-0.7em]
SimLM (baseline)           & 0.185 \drop{(-55.0\%)}     & Dense                       \\
\bottomrule
\end{tabular}
\caption{MRR@10 performance on MSMARCO Dev queries where BM25 fails to retrieve the gold passage within the top-1000 results. Relative performance drops are shown in parentheses.}
\label{tab:robustness_test}
\vskip -0.22in
\end{table}

\subsection{Robustness Analysis}
\label{sec:Robustness-Analysis}
CL‑SR computes query–document similarity not at the lexical term level, but in a semantically generalized latent concept space. Each latent concept is a learned representation of baseline model that can cluster lexically diverse yet semantically related expressions into a single discrete unit. This design preserves the semantic expressiveness, allowing CL‑SR to remain robust to vocabulary mismatch and semantic mismatch.
\subsubsection{Robustness on vocabulary/semantic mismatch}
To empirically verify this robustness, we construct a \textit{Mismatch Set} from the MS MARCO dataset by selecting only the queries for which BM25 fails to retrieve the gold passage within top-1000 results among 8.84 million candidate passages. This subset comprises 988 queries out of the total 6,980 Dev queries, representing failure cases where traditional term-based sparse retrieval is likely to fail due to lexical or semantic discrepancies between queries and relevant documents\footnote{Additional evaluations based on failure cases defined at top-10 and top-100 are reported in Appendix~\ref{sec:APPENDIX_robustness_test}.}.

\begin{figure*}[t]  % [t]: 페이지 상단에 배치
  \centering
  \includegraphics[width=\textwidth]{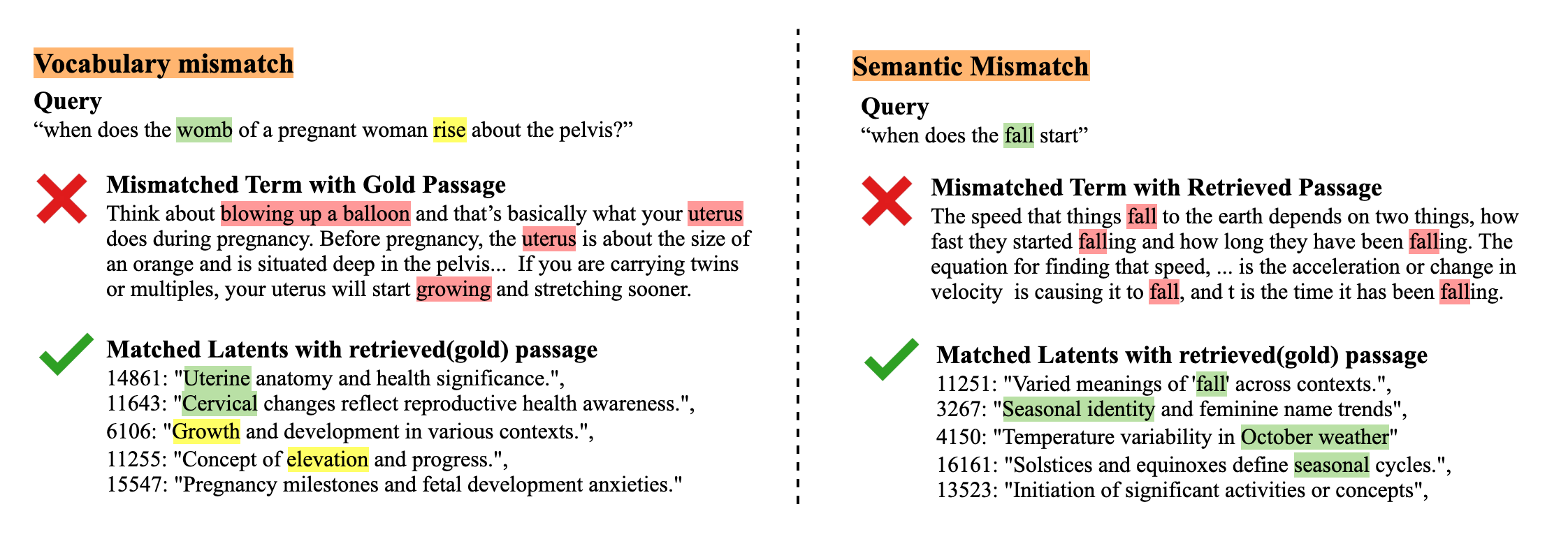}
  \caption{Qualitative case study illustrating CL-SR’s ability to handle vocabulary and semantic mismatch.}
  \label{fig:case_study}
  \vskip -0.2in
\end{figure*}

Experimental results (Table \ref{tab:robustness_test}) show that while traditional sparse models exhibit a significant performance drop on the Mismatch Set, CL‑SR effectiveness remains relatively stable. Notably, SAE-Efficient demonstrates stronger robustness compared to SPLADEv2, and SAE-Max achieves the highest robustness among sparse models and is on par with its target model (SimLM). These findings suggest that latent concepts operate as a semantically rich and generalizable retrieval unit, capable of bridging vocabulary and semantic gaps without relying on query or document expansion mechanisms. 

\subsubsection{Case study}
To further investigate how CL‑SR addresses mismatch scenarios in practice, we conduct a qualitative analysis on the Mismatch Set. Figure \ref{fig:case_study} provides case-studies that illustrate how CL‑SR effectively handles both vocabulary mismatch and semantic mismatch conditions.

\textbf{Vocabulary Mismatch.} In Figure \ref{fig:case_study} left, the query contains the term “womb”, while the corresponding gold passage uses “uterus”. Moreover, the term “rise” in the query is expressed metaphorically in the passage as “blowing up a balloon”. These lexical variations caused BM25 to fail due to the absence of exact term overlap. In contrast, CL‑SR captures mutually activating concepts such as latent 14861 (Uterine anatomy and health significance), 11643 (Cervical changes), and 6106 (Growth and development). These shared activations enable CL‑SR to retrieve the correct passage by aligning abstract concepts, rather than relying on lexical overlap or external term expansion.

\textbf{Semantic Mismatch.} In Figure \ref{fig:case_study} right, the query involves the term “fall”, which can be pragmatically inferred to mean “autumn” from the query itself. However, BM25 assigns high relevance score to an unrelated document where “fall” appears multiple times but in the physical sense (i.e., to drop). CL‑SR overcomes this by distributing semantic meaning across multiple latent concepts, such as latent 11251 (Varied meanings of “fall”) and 3267 (Seasonal identity), thereby capturing the intended sense of the query and retrieving the correct document.

These case studies highlight CL‑SR’s ability to generalize beyond exact lexical term matching by leveraging contextual and conceptual representations, overcoming structural limitations inherent in term-based retrieval frameworks.

\section{Conclusion}
This study proposes a novel interpretability framework for DPR models by leveraging SAEs to decompose dense embeddings—previously considered uninterpretable—into semantically distint latent concepts. Through extensive experiments, we demonstrate that the latent concepts effectively preserve the information contained in the original embeddings while functioning as interpretable semantic units. Building on this, we empirically validate the explainability of both the DPR models embeddings and the similarity scoring process. Additionally, we introduced CL-SR, a novel retrieval paradigm that integrates the semantic expressiveness of dense retrieval with the efficiency and interpretability of sparse retrieval. The proposed CL-SR maintains retrieval accuracy comparable to traditional term-based sparse methods while achieving superior efficiency in terms of both computational cost and storage requirements. Notably, it exhibits strong robustness in challenging scenarios involving vocabulary and semantic mismatches. We believe our framework  provides a new method to gain insights on dense passage retrieval process that can be used to improve dense retrieval or be leveraged to enhance neural sparse retrievals.

\section*{Limitations}
\textbf{Known issues of SAE} Though SAEs are the most popular unsupervised decomposition methods in interpretability, they pose substantial practical and conceptual limitations\citep{sharkey2025open}. Notably, features found by SAEs are known to be incomplete\citep{leask2025sparse, templeton2024scaling, Robert_AIZI_report} and dataset dependent\citep{Kissane_dataset_dependant} and we believe these problems to be present in our SAE as well. Although recent studies try to address these issues by modifying SAE architectures, we stick to the widely used BatchTopK SAE and leave application of newer variants to future work.

\textbf{Evaluation of XAI} Due to the lack of established baselines for explaining ranking models, we primarily rely on LLMs for the evaluation of explanations by instructing them to perform proxy tasks. Though utilizing LLMs for evaluation is widely utilized in SAE literatures, their sensitivity to prompts and variability in reasoning limits the reliability of the evaluation. To compensate, we conduct a human study designed to simulate the ranking model's predictions. This evaluation method requires human annotators and is therefore labor-intensive and difficult to scale, which limited us to evaluating a sampled subset of the data.

\section*{Acknowledgments}
This work was partly supported by the National Research Foundation of Korea (NRF) grant funded by the Korea government (MSIT) (No. RS-2024-00350379, 30), Institute of Information \& Communications Technology Planning \& Evaluation (IITP) grant funded by the Korea government (MSIT) (No. 2022-0-00369, RS-2022-II220369, 30), Institute of Information \& communications Technology Planning \& Evaluation(IITP) grant funded by the Korea government(MSIT) (RS-2019-II190421, Artificial Intelligence Graduate School Program(Sungkyunkwan University),20), and  ICT Creative Consilience Program through the Institute of Information \& Communications Technology Planning \& Evaluation(IITP) grant funded by the Korea government(MSIT) (RS-2020-II201821, 20).

\section*{License}

This work utilizes the \textit{Pyserini} toolkit~\cite{lin2021pyserini}, an open-source Python framework for reproducible information retrieval research with sparse and dense representations, released under the Apache License 2.0.

We also make use of the \textit{MS MARCO} datasets, provided by Microsoft for non-commercial research purposes only. The dataset is distributed ``as is'' and subject to Microsoft’s Terms and Conditions.\footnote{\url{https://microsoft.github.io/msmarco/}}

% References
\bibliography{custom}
\appendix
\section{SAE training Details}
\label{sec:training_details}
We trained our SAE on the MSMARCO corpus and its training queries, which were preprocessed into input sequences of maximum 144 and 32 tokens each for input into our baseline model. Training was performed with a learning rate of $5 \times 10^{-5}$, a batch size of 4096, and for 100 epochs. Training a single SAE model in this setup took about 350 minutes on a single RTX 3090 GPU.

All models were trained using the AdamW \citep{loshchilov2018decoupled} optimizer with $\beta_1 = 0.9$, $\beta_2 = 0.999$, and $\epsilon = 6 \times 10^{-10}$. At each training step, the decoder weights are normalized to 1 following \citet{bricken2023monosemanticity}.

We follow \citet{gao2025scaling}’s method for the auxiliary loss. The total loss is defined as:
\[
\mathcal{L}_{\text{total}} = \mathcal{L} + \lambda \mathcal{L}_{\text{aux}}, \quad \text{where } \lambda = 0.0625.
\]
We consider neurons inactive for 20 training steps as dead and use top 2*k dead neurons for the dead reconstruction.

\section{Experimental Details}
\label{sec:experimental_details}
In this appendix, we provide details about human evaluation, dataset, sampling, LLM prompts of our experiments.

\subsection{Description generation details}
\label{sec:llm_prompt_template}
For BatchTopK SAE, at inference time, latents with activation value larger than mean of top k-th activation over the whole dataset is considered “activated”. For each latent that has been activated at least once over the whole MSMARCO corpus, we provide top 30 most activating MSMARCO passage and instruct GPT4.1-mini to generate description for the latent. We use the prompt of Figure \ref{fig:prompt_templates} (b) for the instruction.

\begin{figure}
 \includegraphics[width=\columnwidth]{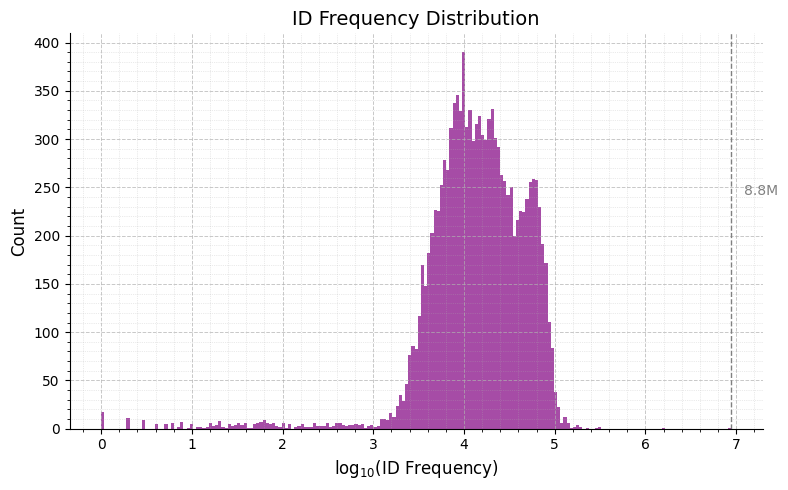}
\caption{Frequency distribution of SAE latents across documents.}
 \label{fig:latent_frequency_distribution}
\end{figure}

\begin{figure}[t]
  \centering
  % (a) latent intrusion
  \begin{subfigure}{0.45\textwidth}
    \begin{PromptBox}
You are an expert linguist analyzing pieces of documents. Below, you will see a set of documents that has some common features, but one of them is an intruder (it does not have that common feature in it).\\[1ex]
Your task is to identify the intruder document and explain why it does not fit.\\[1ex]
The last line of your response must be the formatted response, using \verb|"[intruder]:Document#|"\\\\
\verb|<\nDocument{i} : {passage}>|\\\\
Which document is the intruder, and why?
    \end{PromptBox}
    \subcaption{Prompt template for latent intrusion test.}
    \label{fig:intrusion_template}
  \end{subfigure}

  \vspace{1em}

  % (b) description generation
  \begin{subfigure}{0.45\textwidth}
    \begin{PromptBox}
You are a meticulous AI researcher conducting an important investigation into patterns found in language. Your task is to analyze text and provide an interpretation that thoroughly encapsulates possible patterns found in it.\\[1ex]
\textbf{Guidelines:}\\
You will be given a list of text examples on which a certain common pattern might be present. How important each text is for the pattern is listed after each text.\\[1ex]
- Try to produce a concise final description. Simply describe the text latents that are common in the examples, and what patterns you found.\\
- If the examples are uninformative, you don’t need to mention them. Don’t focus on giving examples of important tokens, but try to summarize the patterns found in the examples.\\
- Based on the found patterns, summarize your interpretation in 1–8 words.\\
- Do not make lists of possible interpretations. Keep your interpretations short and concise.\\
- The last line of your response must be the formatted interpretation, using \verb|[interpretation]:|\\[1ex]
\verb|<Example{i}: {passage} Activation: {act}>|
    \end{PromptBox}
    \subcaption{Prompt template for generating natural‐language descriptions of each latent concept.}
    \label{fig:generation_prompt}
  \end{subfigure}

  \caption{LLM prompt templates for our two tasks: (a) latent intrusion test and (b) description generation.}
  \label{fig:prompt_templates}
\end{figure}

\begin{table}[t]
\centering
\small
\begin{adjustbox}{width=\columnwidth}
\begin{tabular}{lccc}
\toprule
\textbf{Dataset} & \textbf{R@P vs R@P} & \textbf{R@P vs NR@P} & \textbf{NR@P vs R@N} \\
\midrule
\textbf{TREC DL 2019} & 104{,}529 & 196{,}110 & 1{,}004{,}471 \\
\textbf{TREC DL 2020} & 63{,}035  & 84{,}767  & 1{,}099{,}965 \\
\bottomrule
\end{tabular}
\end{adjustbox}
\caption{Document pair statistics for ranking interpretation on TREC DL 2019 and 2020.}
\label{tab:rank_pair_statistics}
\vskip -0.1in
\end{table}

\begin{table*}[t]
\centering
\small
\newcommand{\ratio}[1]{\scalebox{0.7}{\((\times #1)\)}}
\newcommand{\std}[1]{\scalebox{0.7}{\(\pm #1\)}}
\begin{adjustbox}{width=\textwidth}
\begin{tabular}{l c ccc ccc}
\toprule
\multirow{2}{*}{\textbf{Model}} & \multirow{2}{*}{\textbf{NMSE}} &
\multicolumn{3}{c}{\textbf{TREC DL 2019}} &
\multicolumn{3}{c}{\textbf{TREC DL 2020}} \\
\cmidrule(lr){3-5}\cmidrule(lr){6-8}
 & & \textbf{NDCG@10} & \textbf{Recall@1k} & \textbf{Spearman}
   & \textbf{NDCG@10} & \textbf{Recall@1k} & \textbf{Spearman} \\
\midrule
\textbf{Baseline (TAS-B)}
  & --   & 0.721 & 0.783 & --
        & 0.685 & 0.800 & -- \\

Reconstructed (k=32)
  & 0.2106 & 0.573 \ratio{0.80} & 0.685 \ratio{0.88} & 0.865 \std{3.4E-2}
        & 0.517 \ratio{0.76} & 0.676 \ratio{0.85} & 0.861 \std{3.6E-2} \\

Reconstructed (k=48)
  & 0.2005 & 0.575 \ratio{0.80} & 0.690 \ratio{0.88} & 0.894 \std{2.1E-2}
        & 0.523 \ratio{0.76} & 0.700 \ratio{0.88} & 0.891 \std{2.1E-2} \\

Reconstructed (k=64)
  & 0.1610 & 0.593 \ratio{0.82} & 0.695 \ratio{0.89} & 0.928 \std{1.8E-2}
        & 0.565 \ratio{0.83} & 0.716 \ratio{0.90} & 0.925 \std{1.7E-2} \\

Reconstructed (k=128)
  & 0.1188 & \textbf{0.626 \ratio{0.87}} & \textbf{0.735 \ratio{0.94}} & \textbf{0.934 \std{1.4E-2}}
        & \textbf{0.604 \ratio{0.88}} & \textbf{0.753 \ratio{0.94}} & \textbf{0.932 \std{1.4E-2}} \\
\midrule
\textbf{Baseline (GTR-T5)}
  & --   & 0.687 & 0.737 & --
        & 0.664 & 0.721 & -- \\

Reconstructed (k=32)
  & 0.1976 & 0.527 \ratio{0.77} & 0.617 \ratio{0.84} & 0.907 \std{1.4E-2}
        & 0.469 \ratio{0.71} & 0.630 \ratio{0.87} & 0.908 \std{1.2E-2} \\

Reconstructed (k=48)
  & 0.1675 & 0.565 \ratio{0.82} & 0.631 \ratio{0.86} & 0.926 \std{1.1E-2}
        & 0.547 \ratio{0.82} & 0.650 \ratio{0.90} & 0.926 \std{9.3E-3} \\

Reconstructed (k=64)
  & 0.1474 & 0.592 \ratio{0.86} & 0.647 \ratio{0.88} & 0.937 \std{1.0E-2}
        & 0.568 \ratio{0.85} & 0.666 \ratio{0.89} & 0.937 \std{8.3E-3} \\

Reconstructed (k=128)
  & 0.1364 & \textbf{0.590 \ratio{0.86}} & \textbf{0.687 \ratio{0.93}} & \textbf{0.959 \std{7.3E-3}}
        & \textbf{0.614 \ratio{0.93}} & \textbf{0.685 \ratio{0.95}} & \textbf{0.958 \std{6.7E-3}} \\
\bottomrule
\end{tabular}
\end{adjustbox}
\caption{Quantitative evaluation of the SAE trained on TAS-B and GTR-T5 embeddings with the same settings (k={32,48,64,128}) and training data as used for SimLM, demonstrating its generalization performance across different DPR models. Stat. sig. difference w/ paired $t$-test ($p < 0.05$).}

\label{tab:recon_tasb}
\end{table*}

% ---------- Table: Reconstruction & Retrieval Performance on TREC-COVID & NFCorpus ----------
\begin{table*}[t]
\centering
\small
\newcommand{\ratio}[1]{\scalebox{0.7}{\((\times #1)\)}}
\newcommand{\std}[1]{\scalebox{0.7}{\(\pm #1\)}}
\begin{adjustbox}{width=\textwidth}
\begin{tabular}{l ccc ccc}
\toprule
\multirow{2}{*}{\textbf{Model}}
  & \multicolumn{3}{c}{\textbf{TREC-COVID}}
  & \multicolumn{3}{c}{\textbf{NFCorpus}} \\
\cmidrule(lr){2-4}\cmidrule(lr){5-7}
  & \textbf{MRR@10} & \textbf{Recall@1k} & \textbf{Spearman}
  & \textbf{MRR@10} & \textbf{Recall@1k} & \textbf{Spearman} \\
\midrule
\textbf{Baseline (SimLM)}                 
  & 0.828            & 0.237             & --                
  & 0.498            & 0.592             & --                \\

Reconstructed (k=32)
  & 0.692 \ratio{0.83} & 0.188 \ratio{0.79} & 0.887 \std{3.3E\text{-}2}
  & 0.444 \ratio{0.89} & 0.567 \ratio{0.96} & 0.806 \std{2.7E\text{-}2} \\

Reconstructed (k=48)
  & 0.704 \ratio{0.85} & 0.190 \ratio{0.80} & 0.891 \std{3.5E\text{-}2}
  & 0.446 \ratio{0.89} & 0.569 \ratio{0.96} & 0.809 \std{2.9E\text{-}2} \\

Reconstructed (k=64)   
  & 0.728 \ratio{0.88} & 0.193 \ratio{0.82} & 0.914 \std{3.2E\text{-}2}
  & 0.457 \ratio{0.92} & 0.575 \ratio{0.97} & 0.851 \std{6.8E\text{-}2} \\

Reconstructed (k=128)
  & {0.733 \ratio{0.89}} & {0.216 \ratio{0.92}} & {0.930 \std{2.8E\text{-}2}}
  & {0.467 \ratio{0.94}} & {0.571 \ratio{0.96}} & {0.896 \std{2.0E\text{-}2}} \\
\bottomrule
\end{tabular}
\end{adjustbox}
\caption{Zero-shot evaluation of the SAE trained on MSMARCO SimLM embeddings, showing its generalization performance on unseen datasets (TREC-COVID and NFCorpus). Stat. sig. difference w/ paired $t$-test ($p < 0.05$).}
\label{tab:recon_retrieval}
\vskip -0.1in
\end{table*}

\subsection{Dense Embedding Interpretation Details}
\label{sec:dense_emb_interpretability}
For each 600 randomly sampled documents of MSMARCO corpus, we make a set of feature descriptions. Then, for each document, we additionally sample 9 random documents to choose our target document from. We instruct 6 graduate students in machine learning (not authors of this paper) to identify the target passage and report accuracy over 600 samples. We provide a concrete example of the task in Figure \ref{fig:human_template_v1}

\subsection{Ranking Interpretation Details}
\label{sec:ranking_interpretability}
We run SimLM (target model of our SAEs) on TREC 2019, 2020 yielding total combination of document pairs as (Table \ref{tab:rank_pair_statistics}). From each set, we randomly sample 100 pairs, resulting in a total of 600 pairs. We then distribute 600 total prompts to 6 graduate students in machine learning (not authors of this paper) to predict which of the two given documents would have been ranked higher by the target model given latent concepts extracted. We provide a concrete example of the task in Figure \ref{fig:human_template_v2}

\section{Generalizability}
\label{sec:appendix_generalization}

To evaluate the generalizability of the Sparse Autoencoder (SAE) beyond the training distribution, we conduct two sets of transfer experiments, summarized in Table~\ref{tab:recon_tasb} and Table~\ref{tab:recon_retrieval}.

\vspace{2mm}
\noindent\textbf{SAE Generalization Across Baseline DPR Models.} \
Table~\ref{tab:recon_tasb} evaluates the robustness of the SAE when applied to different dense retrievers. Keeping the training settings and MSMARCO data fixed, we replace the SimLM encoder with two off-the-shelf DPR variants—TAS-B\citep{hofstatter2021tasbalanced} and GTR-T5\citep{ni2022gtr}—and retrain the SAE. This allows us to test whether the proposed framework is model-agnostic.

\vspace{2mm}
\noindent\textbf{SAE Transfer to Unseen Datasets.} \
In Table~\ref{tab:recon_retrieval}, We assess whether our SAE trained on MSMARCO passage embeddings from the SimLM model retains effectiveness on unseen datasets.
Since computing Spearman’s correlation over the full ranking is costly, we restrict the evaluation to datasets of manageable size: TREC-COVID (50 queries, 171k passages) and NFCorpus (323 queries, 3.6k passages). 

\section{Impact of IDF Weighting}
\label{sec:APPENIX_IDF_Weighting}

Figure~\ref{fig:latent_frequency_distribution} shows that the SAE(k=32) latents follow a heavy-tail distribution across documents: highly frequent latent concepts are more abstract (and less informative), whereas low-frequency latent concepts capture specific semantics.
\subsection{Interpretability Case Study}
\label{sec:APPENIX_interpretability_case_study}

Figure \ref{fig:IDF_case_study} illustrates a case study on an MSMARCO passage describing cost‐of‐living and inflation benchmarks in the 1960s, showing a subset of latent concepts extracted from the passage’s dense embedding. When no IDF weighting is applied (W/o IDF), high‐frequency, semantically abstract latents—e.g. latent 8033 (“Definitions and concept distinctions in text”)—dominate the activation ranking and obscure more discriminative features. After applying IDF weighting (W/ IDF), these common latents are weakened and truly informative concepts emerge: latent 8983 (“Historical financial and social benchmarks circa 1960”) rises to the top, and latents 21047 (“Income‐related financial and economic concepts”) and 20216 (“1960s–70s cultural and fashion evolution”) receive substantially higher activation ranks compared to the unweighted case, thereby enhancing the interpretability of the embedding.

\begin{figure*}[t] 
  \centering
  \includegraphics[width=\textwidth]{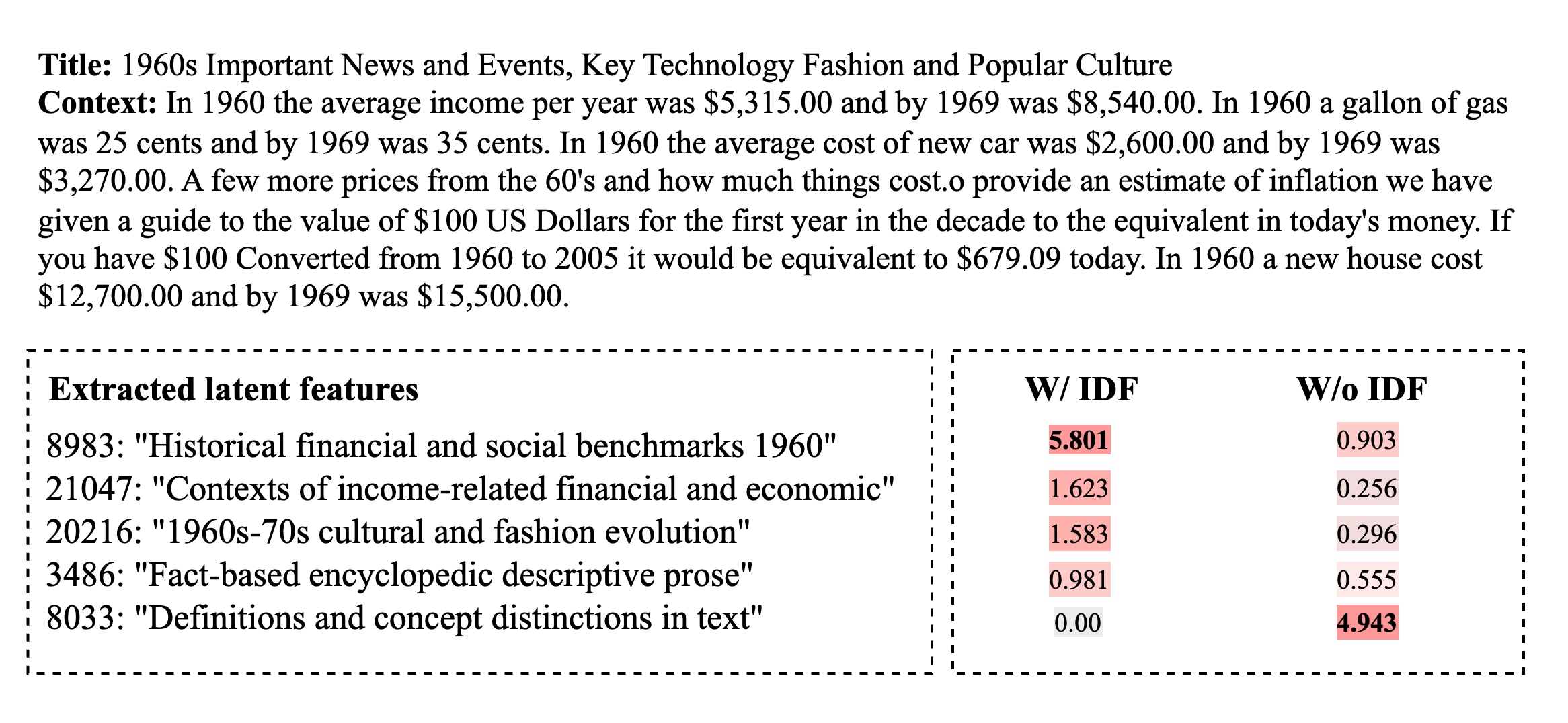}
  \caption{Impact of IDF reweighting on an MSMARCO passage—suppressing abstract latents (8033, 3486) and elevating content-specific latents (8983, 21047, 20216).}
  \label{fig:IDF_case_study}
  \vskip -0.2in
\end{figure*}

\subsection{CL-SR Performance}
\label{sec:APPENIX_IDF_CL_SR_performance}
To quantify the effect of applying Inverse Document Frequency (IDF) to latent activations, we compare two scoring variants:
\begin{itemize}
  \item \textbf{w/ IDF}: activations are scaled by their IDF (Eq.~\ref{eq:latent_bm25}), and  
  \item \textbf{w/o IDF (Dot-product)}: raw activation values are used without any weighting.
\end{itemize}
Table~\ref{tab:perf_idf_comparison} reports the relative performance drop of the dot-product variant (shown in gray) compared to CL-SR. The results show that applying IDF weighting consistently improves retrieval performance by reducing the influence of overly frequent latents.

\begin{table}[t]
\centering
\small
% 회색·작은 글씨 감소폭용 매크로
\newcommand{\drop}[1]{\scalebox{0.7}{\textcolor{gray}{#1}}}

\begin{adjustbox}{width=\columnwidth}
\begin{tabular}{lcc}
\toprule
 & \textbf{w/ IDF} & \textbf{w/o IDF(Dot‐product)} \\
\midrule
% TREC‐DL 제목행 (\\ 필수!)
\multicolumn{3}{@{}l@{}}{\textbf{TREC‐DL 2019 \& 2020 NDCG@10}} \\
\cmidrule(lr){1-3}
CL-SR$_{\mathrm{efficient}}$ & 0.643 / 0.593 & 0.530\drop{(-15.61\%)} / 0.494\drop{(-17.94\%)} \\
CL-SR$_{\mathrm{max}}$       & 0.686 / 0.634 & 0.587\drop{(-13.29\%)} / 0.562\drop{(-10.94\%)} \\
\midrule
% MSMARCO 제목행 (\\ 필수!)
\multicolumn{3}{@{}l@{}}{\textbf{MSMARCO Dev MRR@10 \& Recall@1k}} \\
\cmidrule(lr){1-3}
CL-SR$_{\mathrm{efficient}}$ & 0.343 / 0.954 & 0.274\drop{(-19.65\%)} / 0.908\drop{(-4.52\%)} \\
CL-SR$_{\mathrm{max}}$       & 0.368 / 0.969 & 0.316\drop{(-14.13\%)} / 0.934\drop{(-3.61\%)} \\
\bottomrule
\end{tabular}
\end{adjustbox}

\caption{Retrieval performance comparison between IDF-weighted activations and without IDF (Dot-product) activations. Relative performance drops are shown
in parentheses.}
\label{tab:perf_idf_comparison}
\end{table}

\section{Impact of Latent Count on CL-SR Performance}
\label{sec:APPENDIX_Latent_count}
\begin{table}[t]
\centering
\small
\begin{adjustbox}{width=\columnwidth}
\begin{tabular}{lcccc}
\toprule
\# of Latents 
  & \textbf{MSMARCO Dev} 
  & \textbf{TREC DL 2019} 
  & \textbf{TREC DL 2020} 
  & \textbf{Hyperparameters} \\
  & MRR@10 & NDCG@10 & NDCG@10 & $(k_{1},b,k_{2})$ \\
\midrule
32  & 0.343 & 0.643 & 0.594 & 0.6,\,1.75,\,2.5 \\
48  & 0.353 & 0.646 & 0.621 & 0.6,\,1.25,\,2.0 \\
64  & 0.359 & 0.662 & 0.603 & 0.4,\,0.75,\,2.5 \\
128 & 0.368 & 0.686 & 0.634 & 0.2,\,3.0,\,0.5 \\
\bottomrule
\end{tabular}
\end{adjustbox}
\caption{Performance of CL-SR with varying numbers of latent concepts on MSMARCO Dev and TREC-DL (Recall@1k column omitted).}
\label{tab:clsr_latent_counts}
\end{table}
% ---------- Table: Failure Case MRR@10 on BM25 Misses ----------
\begin{table}[t]
\centering
\small
\newcommand{\drop}[1]{\scalebox{0.7}{\textcolor{gray}{#1}}}
\begin{adjustbox}{width=\linewidth}
\begin{tabular}{lcc}
\toprule
\textbf{Models} & \textbf{Failures@10} & \textbf{Failures@100} \\
\midrule
BM25 & 0.0 \drop{(-100\%)} & 0.0 \drop{(-100\%)} \\
docT5query & 0.129 \drop{(-52.4\%)} & 0.08 \drop{(-70.5\%)} \\
DeepImpact & 0.192 \drop{(-41.2\%)} & 0.123 \drop{(-62.4\%)} \\
SPLADE-Max & 0.218 \drop{(-35.8\%)} & 0.149 \drop{(-56.1\%)} \\
CL-SR (Efficient) & 0.226 \drop{(-33.9\%)} & 0.163 \drop{(-52.3\%)} \\
\textbf{CL-SR (Max)} & \textbf{0.248} \drop{(-32.6\%)} & \textbf{0.185} \drop{(-49.7\%)} \\
\bottomrule
\end{tabular}
\end{adjustbox}
\caption{MRR@10 performance on MSMARCO Dev queries where BM25 fails to retrieve the gold passage within the top-$K$ candidates. Relative performance drops are shown in parentheses.}
\label{tab:appendix_failure_robustness}
\vskip -0.22in
\end{table}
We study how the number of latent concepts \(k\) affects both effectiveness and hyperparameter settings of Concept-Level Sparse Retrieval (CL-SR). We vary \(k\) over \{32, 48, 64, 128\}, tuning the BM25-style parameters \((k_{1}, b, k_{2})\) on the MSMARCO Dev set for each configuration. 

Table~\ref{tab:clsr_latent_counts} reports MRR@10 and Recall@1k on MSMARCO Dev, as well as NDCG@10 on TREC-DL 2019/2020. As \(k\) increases from 32 to 128, we observe a steady improvement in all metrics: MRR@10 rises from 0.343 to 0.368, Recall@1k from 0.954 to 0.969, and NDCG@10 on TREC-DL from 0.643/0.593 to 0.686/0.634. These gains confirm that the number of latents used for reconstruction increases, the original embedding information can be recovered more faithfully.

\section{Robustness Test on Mismatch Set}
\label{sec:APPENDIX_robustness_test}
To complement the main analysis based on top-1000 failures, we additionally evaluate retrieval performance on subsets where BM25 fails to retrieve the gold passage at top-10 (4.3k queries) and top-100 (2.3k queries) on MSMARCO DEV queries. Table~\ref{tab:appendix_failure_robustness} reports the results. We observe that CL-SR (Efficient) and CL-SR (Max) consistently exhibit substantially smaller performance drops.

\section{Comparison with Neuron-based Interpretability}
\label{sec:APPENDIX_neuron_comparison}
Previous works like \citet{Zhan2021InterpretingDR} has tried to chunk and interpret vector representations in standard basis. We believe this could be problematic since these ranking models are trained with cosine similarity which is rotation invariant. Our method on the other hand does not assume that dimensions have a spatial meaning and instead learns an interpretable basis. Previous works like \citet{olah2020zoom} also shows that individual neurons (or spatial dimension) are often polysemantic and we believe this holds in our setting as well, motivating our usage of sparse autoencoders.
To empirically test this, we performed the latent intrusion test (as described in ~\ref{sec:Interpreting_DPR}) to the 768-dimensional hidden representation from SimLM. Under the same setting, the accuracy was 0.46, compared to 0.86 with our SAE based method, highlighting a substantial gap in interpretability.

% ---------- Table: Efficiency Comparison on MSMARCO Dev ----------
\begin{table}[t]
\centering
\small
\newcommand{\drop}[1]{\scalebox{0.7}{\textcolor{gray}{#1}}}
\newcommand{\na}{\textemdash}
\begin{adjustbox}{width=\linewidth}
\begin{tabular}{lccccc}
\toprule
\textbf{Model} & \textbf{Index (GB)} & \textbf{Factor} &
\shortstack[c]{\textbf{Encode Latency}\\ \textbf{(GPU, ms/query)}} &
\shortstack[c]{\textbf{Search Latency}\\ \textbf{(CPU, ms/query)}} &
\shortstack[c]{\textbf{Search Latency}\\ \textbf{(GPU, ms/query)}} \\
\midrule
SimLM                 & 26.0 & $\times$8.52 & \textbf{3.42} & 924.42  & \textbf{5.58} \\
SPLADE                & 2.6  & $\times$0.85 & 3.63          & 163.71  & \na \\
CL-SR (Max)           & 1.27 & $\times$0.42 & 3.89          & 154.25  & \na \\
\textbf{CL-SR (Efficient)} & \textbf{0.57} & $\times$0.19 & 3.86          & \textbf{117.96} & \na \\
\bottomrule
\end{tabular}
\end{adjustbox}
\caption{Efficiency comparison on MSMARCO Dev. Lower is better. Bold indicates the best in each column.}
\label{tab:appendix_efficiency_msmarco}
\vskip -0.22in
\end{table}

\section{Efficiency Comparison with Dense Retrieval}
\label{sec:appendix_efficiency_comparison}
To provide a more comprehensive comparison, we additionally report an efficiency analysis against the target dense model, SimLM, in Table~\ref{tab:appendix_efficiency_msmarco}. We provide our experimental results on the MSMARCO Dev set following the implementation of in \citet{li2022citadel} with PyTorch (GPU, Single Nvidia RTX 3090) and Numpy (CPU, Single Intel(R) Core(TM) i7-14700KF CPU @ 3.40 GHz). We set the batch size to 1 and ran all 6,980 MS MARCO Dev queries three times—reporting the lowest observed latency.

\begin{table*}[ht]
  \centering
  \scriptsize
  \setlength{\tabcolsep}{4pt}  % 셀 좌우 여백 조정
  \begin{adjustbox}{width=\textwidth}
    \begin{tabular}{@{}l|p{0.35\textwidth}|p{0.57\textwidth}@{}}
      \toprule
      \textbf{Source} & \textbf{Text} & \textbf{Top Activated Latents (ID: description (activation))} \\
      \midrule
      \textbf{Query} &
      which pathogen depends on living cells &
      \begin{minipage}[t]{\linewidth}
        8327: Pathogens causing disease in humans via blood (5.1931)\\
        8442: Immune system influenced by stress and lifestyle (3.3578)\\
        11271: Living as active state or ongoing condition (3.2910)\\
        4166: Definitions of concepts involving multiple components or functions (2.0142)\\
        8190: Response as reaction, answer, or measured outcome (1.9798)\\
        3580: Multifaceted encyclopedic contexts for “Blood” (1.7898)\\
        1322: Defining unicellular organisms by single-cell simplicity (1.6423)
      \end{minipage}
      \\[4pt]
      \midrule
      \makecell[l]{\textbf{Retrieved }\\\textbf{Incorrect Passage}} &
      The ability of a multicellular organism to defend itself against invasion by pathogens (bacteria, fungi, viruses, etc.) depends on its ability to mount immune responses. &
      \begin{minipage}[t]{\linewidth}
        8306: Definitions of concepts involving multiple components or functions (3.8576)\\
        8442: Immune system influenced by stress and lifestyle (2.9287)\\
        1322: Defining unicellular organisms by single-cell simplicity (2.9263)\\
        8327: Pathogens causing disease in humans via blood (2.8417)\\
        10630: Multifaceted protective actions across legal, psychological, and social domains (1.9795)\\
        6445: Polysemous term "host" across diverse domains (1.9598)\\
        10951: Ability as actualized skill versus potential capacity (1.8653)
      \end{minipage}
      \\[4pt]
      \midrule
      \makecell[l]{\textbf{Not Retrieved}\\\textbf{Gold Passage}} &
      Yes, viruses need a living host to replicate. A virion needs to be inside a living cell in order to hijack that cell so that more virus particles can be made by the cell. Since a virus is not a living thing, it doesn't reproduce in the same way. &
      \begin{minipage}[t]{\linewidth}
        10944: Viruses as pseudo-living entities with classification challenges (5.6177)\\
        11271: Living as active state or ongoing condition (4.7194)\\
        8327: Pathogens causing disease in humans via blood (3.6580)\\
        6445: Polysemous term "host" across diverse domains (2.7850)\\
        256: DNA semiconservative replication mechanism (2.3768)\\
        15511: Single-parent reproduction, genetic uniformity, rapid population growth (2.3407)\\
        8296: Immune cells engulfing and digesting foreign material (1.6999)
      \end{minipage}
      \\
      \bottomrule
    \end{tabular}
  \end{adjustbox}
  \caption{Latent concept activations for a failed query from the MS MARCO Dev set (i.e.\ R@1000 failure), top-ranked (incorrect) passages, and the gold passage. Each latent concept is listed with its idf-scaled activation value in parentheses.}
  \label{tab:dpr_interpretability_casestudy}
\end{table*}

\section{Interpretability Case Study}
\label{sec:APPENDIX_DPR_Interpretability}
Table~\ref{tab:dpr_interpretability_casestudy} illustrates SAE-based decomposition of SimLM’s dense embeddings for the query “which pathogen depends on living cells,” one of the top-ranked incorrect passage, and the non retrieved gold passage. All three embeddings activate latent 8327 (“pathogens causing disease in humans via blood”), capturing the generic pathogen concept. However, the retrieved (incorrect) passage uniquely shares immune-system latent 8442 (“immune system influenced by stress and lifestyle”) with the query, boosting its matching score to be top-ranked. In contrast, the gold passage activates latent 10944 (“viruses as pseudo-living entities requiring host cells”), which encode the virus-specific host-dependency semantics needed to answer the query; because these latents do not co-occur with the query embedding, SimLM underestimates the passage's relevance. This case study demonstrates that our SAE-based framework can provide a transparent, concept-level diagnosis of retrieval failures.

\begin{figure*}[t] 
  \centering
  \includegraphics[width=\textwidth]{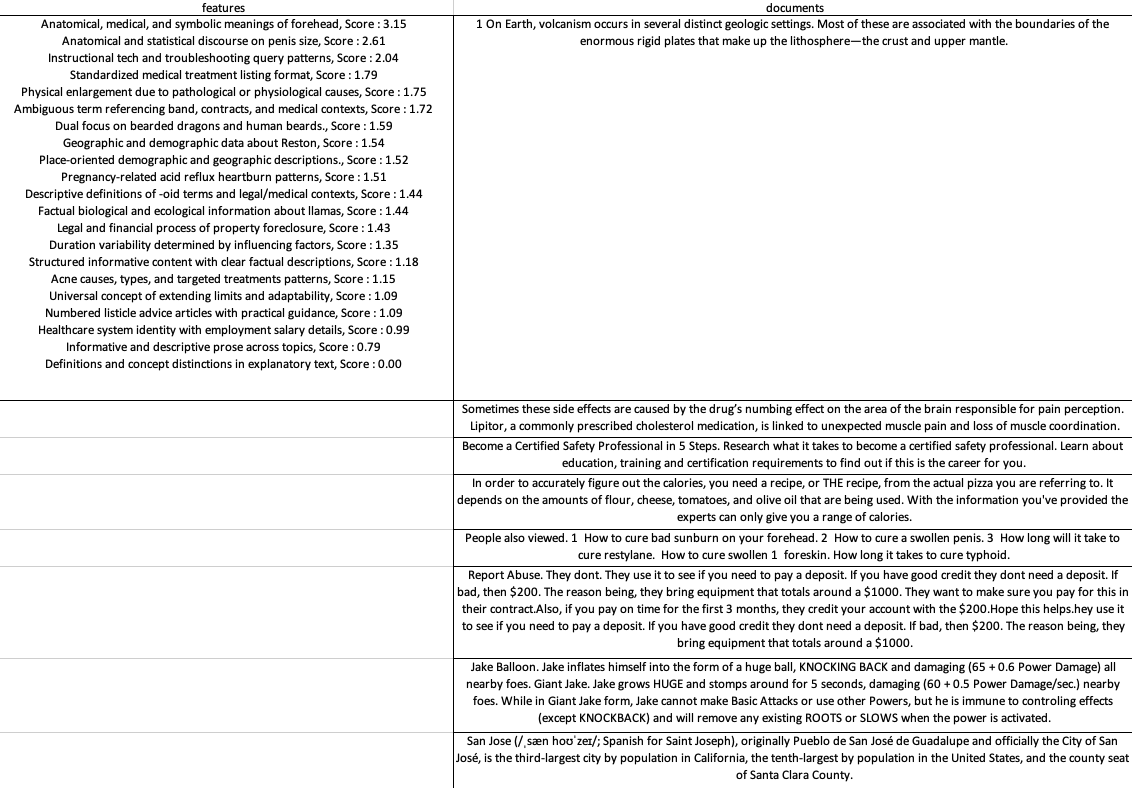}
  \caption{
Example of human evaluation for dense embeddding interpretability. The left column lists the top-activated latent concepts extracted from a single document, along with their descriptions and activation scores. The right column shows the candidate documents. List of documents are truncated in this figure.
}
  \label{fig:human_template_v1}
  \vskip -0.2in
\end{figure*}

\begin{figure*}[t] 
  \centering
  \includegraphics[width=\textwidth]{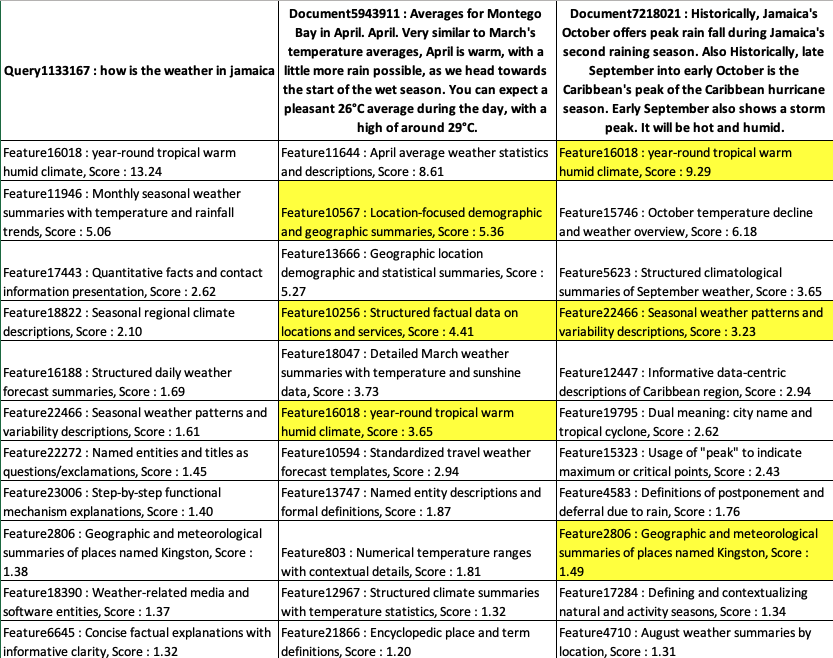}
  \caption{Example of human evaluation for ranking interpretability. The first column shows the query and its extracted latent features, while the second and third columns list latent features extracted from two candidate documents. Features activated by both the query and each document are highlighted. List of features are truncated in this figure.
}
  \label{fig:human_template_v2}
  \vskip -0.2in
\end{figure*}

\end{document}